\RequirePackage[2020-02-02]{latexrelease}
\documentclass[preprintnumbers,nofootinbib,noshowpacs,eqsecnum,prd,superscriptaddress,letterpaper]{revtex4}
\pdfoutput=1
\usepackage{graphicx}
\usepackage{amsmath,amssymb}
\usepackage{url}
\usepackage{multirow}
\usepackage{hyperref,color}
\usepackage[normalem]{ulem}
\usepackage{slashed}
\usepackage{rotating}
\usepackage{afterpage}
\usepackage{ifthen}
\usepackage{mciteplus}
\usepackage{multirow}
\usepackage{color}

\newcommand{\ocal}{{\cal O}}

\newcommand{\sla}[1]{/\!\!\!#1}

\def\lsim{\raise0.3ex\hbox{$\;<$\kern-0.75em\raise-1.1ex\hbox{$\sim\;$}}}
\def\gsim{\raise0.3ex\hbox{$\;>$\kern-0.75em\raise-1.1ex\hbox{$\sim\;$}}}

\newcommand {\obw}     {{\cal O}_{BW}}
\newcommand {\opone}   {{\cal O}_{\Phi,1}}
\newcommand {\owww}    {{\cal O}_{WWW}}
\newcommand {\ow}      {{\cal O}_{W}}
\newcommand {\ob}      {{\cal O}_{B}}
\newcommand {\oww}     {{\cal O}_{WW}}
\newcommand {\obb}     {{\cal O}_{BB}}
\newcommand {\ogg}     {{\cal O}_{GG}}
\newcommand {\optwo}   {{\cal O}_{\Phi,2}}

\newcommand{\eh}{\hat{e}}
\newcommand{\sh}{\hat{s}}
\newcommand{\ch}{\hat{c}}
\newcommand{\vh}{\hat{v}}
\newcommand{\tf}{\tilde{f}}
\newcommand{\tfw}{\tilde{f}_W}
\newcommand{\tfb}{\tilde{f}_B}

\begin{document}

\preprint{YITP-SB-2023-04, FERMILAB-PUB-23-134-V,UWThPh 2023-13}

\title{Impact of dimension-eight SMEFT operators in the EWPO and Triple Gauge Couplings analysis in Universal SMEFT}

\author{Tyler Corbett}
\email{corbett.t.s@gmail.com}
\affiliation{Faculty of Physics, University of Vienna, Boltzmanngasse 5, A-1090 Wien, Austria}
\author{Jay Desai}
\email{jay.desai@stonybrook.edu}
\affiliation{C.N. Yang Institute for Theoretical Physics,
  Stony Brook University, Stony Brook New York 11794-3849, USA}
\author{O.\ J.\ P.\ \'Eboli}
\email{eboli@if.usp.br}
\affiliation{Instituto de F\'{\i}sica, 
Universidade de S\~ao Paulo, S\~ao Paulo -- SP, Brazil.}
\affiliation{Departament de Fis\'{\i}ca Qu\`antica i
  Astrof\'{\i}sica and Institut de Ciencies del Cosmos, Universitat de
  Barcelona, Diagonal 647, E-08028 Barcelona, Spain}
\author{M.~C.~Gonzalez-Garcia}
\email{maria.gonzalez-garcia@stonybrook.edu}
\affiliation{C.N. Yang Institute for Theoretical Physics,
  Stony Brook University, Stony Brook New York 11794-3849, USA}
\affiliation{Departament de Fis\'{\i}ca Qu\`antica i
  Astrof\'{\i}sica and Institut de Ciencies del Cosmos, Universitat de
  Barcelona, Diagonal 647, E-08028 Barcelona, Spain}
\affiliation{Instituci\'o Catalana de Recerca i Estudis
  Avan\c{c}ats (ICREA), Pg. Lluis Companys 23, 08010 Barcelona,
  Spain.}
\author{Matheus Martines}
\email{matheus.martines.silva@usp.br}
\affiliation{Instituto de F\'{\i}sica, 
Universidade de S\~ao Paulo, S\~ao Paulo -- SP, Brazil.}
\author{Peter Reimitz}
\email{peter@if.usp.br}
\affiliation{Instituto de F\'{\i}sica, 
Universidade de S\~ao Paulo, S\~ao Paulo -- SP, Brazil.}
\affiliation{Particle Theory Department, Fermilab, P.O. Box 500, Batavia, IL 60510, USA}

\begin{abstract}We perform a complete study of the electroweak precision
  observables and electroweak gauge boson pair production in terms of
  the SMEFT up to ${\cal O}(1/\Lambda^4)$ under the assumption of
  universal, C and P conserving new physics.  We show that the
  analysis of data from those two sectors allows us to obtain closed
  constraints in the relevant parameter space in this scenario.  In
  particular we find that the Large Hadron Collider data can
  independently constrain the Wilson coefficients of the dimension-six
  and -eight operators directly contributing to the triple gauge boson
  vertices. Our results show that the impact of dimension-eight
  operators in the study of triple gauge couplings is small.
\end{abstract}

\maketitle

\section{Introduction}

During the last decade the Large Hadron Collider (LHC) has accumulated
a large amount of data that lead to further tests of the Standard
Model (SM) and the search for Physics beyond the Standard Model
(BSM). Presently there is no smoking gun indication of any extension
of the SM. Therefore, one can assume that there is a mass gap between
the electroweak scale and the BSM scale. In this scenario, the use of
Effective Field Theory (EFT)~\cite{Weinberg:1978kz, Georgi:1985kw,
  Donoghue:1992dd} as the tool to search for hints of new Physics has
become customary. \medskip

The EFT approach is suited for model--independent analyses since it is
based exclusively on the low-energy accessible states and symmetries.
Assuming that the scalar particle observed in 2012~\cite{ Aad:2012tfa,
  Chatrchyan:2012xdj} belongs to an electroweak doublet, we can
realize the $SU(2)_L \otimes U(1)_Y$ symmetry linearly. The resulting
model is the so-called Standard Model EFT (SMEFT).  There have been
many analyses of the LHC data using dimension-six SMEFT; see for
instance ~\cite{ Corbett:2012dm, Corbett:2012ja, Ellis:2014jta,
  Corbett:2015ksa, Butter:2016cvz, Baglio:2017bfe,
  Aguilar-Saavedra:2018ksv, Ellis:2018gqa, daSilvaAlmeida:2018iqo,
  Brivio:2019ius, Ellis:2020unq, Dawson:2020oco, Ethier:2021bye,
  Almeida:2021asy} and references therein. In order to access the
convergence of the $1/\Lambda$ expansion, as well as avoid the
appearance of phase space regions where the cross section is
negative~\cite{Baglio:2017bfe}, it is important to perform the full
calculation at order $1/\Lambda^4$. The consistent calculation at
order $1/\Lambda^4$ requires the introduction of the contributions
stemming from dimension-eight operators. In the most general scenario,
the number of dimension-eight operators contributing to the present
observables is extremely large ~\cite{Henning:2015alf} and that
precludes a complete general analysis including all effects up to
order $1/\Lambda^4$. Due to its complexity, the systematic study of
the $\ocal(1/\Lambda^4)$ effects is still in its early stages.  To
date there have been a few case studies for
Drell-Yan~\cite{Alioli:2020kez,Boughezal:2021tih,
  Boughezal:2022nof, Kim:2022amu, Allwicher:2022gkm}, $\bar{t}tH$
production~\cite{Dawson:2021xei}, the production of electroweak gauge
boson pairs~\cite{Degrande:2023iob}, Higgs boson
processes~\cite{Hays:2018zze, Corbett:2021cil,Martin:2021cvs,Corbett:2021jox,Hays:2020scx}, and the electroweak
precision data~\cite{ Hays:2020scx, Corbett:2021eux}. \medskip

With this motivation, we perform a complete study of the electroweak
precision observables (EWPO) and electroweak diboson (EWDB) production
at order $1/\Lambda^4$ including all relevant dimension-six and
dimension-eight operators {\sl under the assumption of universal New
  Physics with conservation of $C$ and $P$}~\cite{Wells:2015uba} so
that the EFT contains only bosonic operators after field
redefinitions. In this case, we show that the analysis of existing
data from those two sectors allows one to obtain closed constraints on
the the full relevant parameter space.  Furthermore, we argue that it
is still possible to perform the analysis sequentially, obtaining
first the constraints on four effective combinations of Wilson
coefficients using the EWPO, and then apply those bounds to reduce the
number of Wilson coefficients which are relevant for the the diboson
analysis. Besides demonstrating the feasibility of the analysis and
deriving the corresponding bounds, our main result is to show that in
this scenario the impact of dimension-eight operators in our present
determination of the triple gauge couplings (TGC) is small. \medskip

This work is organized as follows .  The analysis framework employed
is presented in Sec.~\ref{sec:model}.
Sections~\ref{sec:ewpo}~and~\ref{sec:results} contain the results of
the analysis of EWPO and of the EWDB data respectively.  In
Sec.~\ref{sec:disc} we summarize our conclusions.  We present in the
appendices the full expressions of the couplings of the electroweak
gauge bosons to fermions and TGC to order $\ocal(1/\Lambda^4)$ in this
scenario. \medskip

\section{Analysis framework}
\label{sec:model}

Following~\cite{Wells:2015uba}, we consider a theory as universal if
its EFT can be expressed exclusively in terms of bosonic operators via
field redefinitions. We will also assume conservation of $C$ and $P$.
The requirement of the EFT to be universal limits the number of
operators that have to be considered and in Ref.~\cite{Wells:2015uba}
the independent set of dimension-six operators for universal theories
is explicitly worked out in several bases. In this work we use the
Hagiwara, Ishihara, Szalapski, and Zeppenfeld (HISZ) dimension-six
basis~\cite{Hagiwara:1993ck, Hagiwara:1996kf}.
The relevant set of operators left in HISZ basis for universal
theories can be straightforwardly adapted from the results in
Ref.~\cite{Wells:2015uba} for the SILH basis \cite{Giudice:2007fh}
taking into account the different choice of two of the bosonic
operators left in the basis.  With this, one finds that in the HISZ
basis universal theories are described by 11 bosonic operators and 5
fermionic operators. The 11 bosonic operators are:
\begin{alignat}{3}
&\opone =(D_\mu\Phi)^\dagger\Phi\Phi^\dagger(D^\mu\Phi)\;,
&&
\optwo =\frac{1}{2} \partial^\mu\left ( \Phi^\dagger \Phi \right)
\partial_\mu\left ( \Phi^\dagger \Phi
\right ) \;,&&
\hspace*{0.2cm}
\ocal_{\Phi^6} = (\Phi^\dagger \Phi)^3  \;,\;
 \nonumber\\
  & \oww  = \Phi^{\dagger} \widehat{W}_{\mu \nu} \widehat{W}^{\mu \nu} \Phi
  \;, &\hspace*{0.5cm}
  & \obb = \Phi^{\dagger} \widehat{B}_{\mu \nu} \widehat{B}^{\mu \nu}
       \Phi \;,&\hspace*{0.3cm}&\obw =
       \Phi^\dagger\widehat{B}_{\mu\nu}\widehat{W}^{\mu\nu}\Phi  \;,\;
        \nonumber
     \\
  &\ow =    (D_\mu\Phi)^\dagger\widehat{W}^{\mu\nu}(D_\nu\Phi) \;,&&
     \ob =      (D_\mu\Phi)^\dagger\widehat{B}^{\mu\nu}(D_\nu\Phi)
     \;, && \owww=  {\rm
       Tr}[\widehat{W}_{\mu}^{\nu}\widehat{W}_{\nu}^{\rho}\widehat{W}_{\rho}^{\mu}]  \;,\;
     \nonumber
  \\
   &     \ogg  = \Phi^\dagger \Phi \; G_{\mu\nu}^a G^{a\mu\nu}
      \;, && \ocal_{GGG} = g_s^3 f^{abc}{G}_{\mu}^{a\,\nu}
        {G}_{\nu}^{b\,\rho} {G}_{\rho}^{c\,\mu}  \;, &&
\label{eq:dim6-uni}
\end{alignat}
 where $\Phi$ stands for the SM Higgs doublet and we have defined
 $\widehat{B}_{\mu\nu} \equiv i(g^\prime/2)B_{\mu\nu}$ and
 $\widehat{W}_{\mu\nu} \equiv i(g/2)\sigma^aW^a_{\mu\nu}$, with $g_s$,
 $g$ and $g^\prime$ being the $SU(3)_C$, $SU(2)_L$ and $U(1)_Y$ gauge
 couplings, respectively. $\sigma^a$ stands for the Pauli matrices
 while $f^{abc}$ are the $SU(3)_C$ structure constants. \medskip

 Five four-fermion operators are generated when applying the equations
 of motion (EOM) to eliminate bosonic operators involving the square
 of derivatives of the gauge strength tensors and four Higgs fields in
 total analogy with the SILH basis in Ref.~\cite{Wells:2015uba}:
\begin{alignat}{2}
  & \ocal_y = | \Phi|^2 ( \Phi_\alpha J_y^\alpha + {\rm h.c.} ) \;,
    && \ocal_{2y} =  J^\dagger_{y\alpha} J^\alpha_y  \;,
\nonumber   \\
  &\ocal_{2JW} = \sum_{f,f^\prime \in \{Q,L\}} \left( \bar{f} \gamma_\mu
  \frac{\sigma^a}{2} f \right) \left( \bar{f^\prime} \gamma^\mu
    \frac{\sigma^a}{2} f^\prime \right) \;,
     &\hspace*{0.4cm}&   \ocal_{2JB} = \sum_{f,f^\prime \in \{Q,L,u,d,e \}} \left( Y_f
       \bar{f}
           \gamma_\mu f \right) \left( Y_{f^\prime} \bar{f^\prime}
           \gamma^\mu f^\prime \right) \;,
                                          \nonumber
  \\
  &\ocal_{2JG} = \sum_{f,f^\prime \in \{ Q,u,d\} } \left( \bar{f}
    \gamma_\mu T^a f \right) \left( \bar{f^\prime}
    \gamma^\mu T^a f^\prime \right) \;,
     &&
\end{alignat}
where $Y_f$ are the hypercharges, $Q$ and $L$ are the quark and lepton
doublets and $u$, $d$ and $e$ represent the fermion singlets. In
addition, $T^a$ are the Gell-Mann matrices, $y_f$ are the
Yukawa matrices, and
\begin{equation*}
  J_y^\alpha = \bar{u} y_u^\dagger Q_\beta \epsilon^{\alpha\beta} +
  \bar{Q}^\alpha y_d d + \bar{L}^\alpha y_e e \;.
\end{equation*}

For the dimension-eight operators, we will work in the basis 
defined in Ref.~\cite{Murphy:2020rsh}. For universal theories 
the potentially relevant bosonic operators for
our analyses belong to the classes $\Phi^6D^2$, $X^3\Phi^2$,
$X^2 \Phi^4$, and $X \Phi^4 D^2$ with $X$ standing for a field
strength tensor. This includes:\\
$\bullet$ two operators in the class $\Phi^6 D^2$
related to the dimension-six $\ocal_{\Phi,1}$ and $\ocal_{\Phi,2}$:
\begin{equation}
  \ocal^{(1)}_{D^2 \Phi^6}
  = (\Phi^\dagger \Phi)^2 (D_\mu\Phi)^\dagger
     D^\mu \Phi
  \qquad \hbox{and} \qquad
  \ocal^{(2)}_{D^2\Phi^6} =  (\Phi^\dagger \Phi) (\Phi^\dagger
     \sigma^I \Phi) (D_\mu\Phi)^\dagger \sigma^I
     D^\mu \Phi \;,
\end{equation}\\
$\bullet$ two $CP$ conserving operators in class $X^3 \Phi^2$
that contribute to the EWDB analysis are
\begin{equation}
  \ocal^{(1)}_{W^3\Phi^2} = (\Phi^\dagger \Phi)  {\rm Tr}[\widehat{W}^\mu_\nu
  \widehat{W}^\nu_\rho \widehat{W}^\rho_\mu] \qquad \hbox{and} \qquad
  \ocal^{(1)}_{W^2B\Phi^2} = \frac{g^3 s_W}{8 c_W} \epsilon^{IJK}
      \Phi^\dagger \sigma^I \Phi B^\mu_\nu W^{J\, \nu}_ \rho W^{K \,
        \rho}_ \mu \;,
\end{equation}\\
$\bullet$ two operators in class $X \Phi^4 D^2$ contributing to anomalous
TGC are siblings of dimension-six operators $\ob$ and $\ow$:
\begin{equation}
  \ocal^{(1)}_{B\Phi^4 D^2} = (\Phi^\dagger \Phi) (D_\mu \Phi)^\dagger
  \widehat{B}^{\mu\nu} D_\nu \Phi \qquad \hbox{and} \qquad
  \ocal^{(1)}_{W \Phi^4 D^2} =  (\Phi^\dagger \Phi) (D_\mu \Phi)^\dagger
  \hat{W}^{\mu\nu} D_\nu \Phi \; ,
\end{equation}\\
$\bullet$ four operators in the $ X^2 \Phi^4$ class
\begin{alignat}{2}
  & \ocal^{(1)}_{W^2  \Phi^4} = (\Phi^\dagger \Phi)  \Phi^\dagger \widehat{W}_{\mu\nu}
    \widehat{ W}^{\mu\nu} \Phi  \; ,
    &\qquad& \ocal^{(1)}_{B^2 \Phi^4} = (\Phi^\dagger \Phi)^2 \widehat{B}_{\mu\nu}
      \widehat{ B}^{\mu\nu} \; ,
  \\
  &\ocal^{(1)}_{BW \Phi^4} = (\Phi^\dagger \Phi)  \Phi^\dagger \widehat{W}_{\mu\nu} \Phi
  \widehat{B}^{\mu\nu} \;,
  &\qquad&\ocal^{(3)}_{W^2\Phi^4} =  \Phi^\dagger \widehat{W}_{\mu\nu} \Phi
  \Phi^\dagger \widehat{W}^{\mu\nu} \Phi  \; .
\label{eq:x2phi4}
\end{alignat}

In addition some dimension-eight fermionic operators will be generated
by the EOM in analogy to the dimension-six case. Presently there there
is no study of the fermionic operators compatible with universal
theories for the dimension-eight basis. So in what follows, we assume
that only four-fermion operators are generated in exchanging a subset of 
the purely bosonic operators defining the universal basis for fermionic operators. \medskip

It is important to notice that not all operators listed above appear
in the analysis of EWPO and EWDB data even after accounting for their
finite renormalization contribution to the SM parameters.  In this
work, we adopt as input parameters
$\{ \widehat{\alpha}_{\rm em} \;,\; \widehat{G}_F \;,\;
\widehat{M}_Z\}$ and consider the following three relations to define
the renormalized parameters
\begin{eqnarray}
  &&\hat{e} = \sqrt{ 4 \pi \hat{\alpha}_{\rm em}} \;,
     \nonumber
  \\
  && \hat{v}^2 = \frac{1}{\sqrt{2} \, \hat{G}_F} \;,
     \label{eq:vhat}
  \\
  &&\hat{c}^2 \hat{s}^2 = \frac{\pi \widehat{\alpha}_{\rm em}}{\sqrt{2}\,
     \widehat{G}_F \widehat{M}^2_Z} \;,
     \nonumber
\end{eqnarray}
where $\sh$ ($\ch$) is the sine (cosine) of the weak mixing angle
$\hat{\theta}$. \medskip

The predictions of SMEFT at order $1/\Lambda^4$ and the input
parameters in Eq.~(\ref{eq:vhat}) allow us to obtain the SM mixing
angle, electric charge, and the Higgs vev as a function of the input
parameters and some of dimension-six and -eight Wilson coefficients.
In this process the operators $\oww$, $\obb$,
$\ocal^{(1)}_{W^2\Phi^4}$, and $\ocal^{(1)}_{B^2\Phi^4}$ induce an
overall renormalization of the $W^a$ and $B$ field wave functions that
can be absorbed by a redefinition of the coupling constants.
Furthermore, the contribution of $\ocal^{(1)}_{D^2 \Phi^6}$ to the
Higgs vev cancels against its contribution to the renormalization of
the $W^a$ and $B$ field wave functions.  Consequently their
coefficients drop out of any of the predictions in the EWPO and EWDB data
(see the appendix for the explicit expressions). \medskip

Altogether the effective Lagrangian considered in this work reads:
\begin{eqnarray}
{\cal L}_{\rm eff}  =&& {\cal L}_{SM}
+ \frac{f_{WWW}}{\Lambda^2} {\cal  O}_{WWW}
+ \frac{f_{W}}{\Lambda^2} {\cal O}_{W}
+ \frac{f_{B}}{\Lambda^2} {\cal O}_{B}
+ \frac{f_{BW}}{\Lambda^2} {\cal O}_{BW}
    + \frac{f_{\Phi,1}}{\Lambda^2} {\cal O}_{\Phi,1}
  +     \frac{f_{4F}}{\Lambda^2}   {\cal O}_{4F}  
\nonumber
\\
&&
  +     \frac{f^{(2)}_{D^2\Phi^6 }}{\Lambda^4}   {\cal O}^{(2)}_{D^2 \Phi^6 }
  +     \frac{f^{(1)}_{W^3\Phi^2 }}{\Lambda^4}   {\cal O}^{(1)}_{W^3\Phi^2 }
  +     \frac{f^{(1)}_{W^2 B \Phi^2 }}{\Lambda^4}   {\cal
    O}^{(1)}_{W^2 B \Phi^2 }
  +     \frac{f^{(1)}_{B \Phi^4 D^2}}{\Lambda^4}   {\cal O}^{(1)}_{B\Phi^4 D^2}
\nonumber
  \\
  && +     \frac{f^{(1)}_{W \Phi^4 D^2}}{\Lambda^4}   {\cal O}^{(1)}_{W\Phi^4 D^2}
+     \frac{f^{(3)}_{W^2 \Phi^4}}{\Lambda^4}   {\cal O}^{(3)}_{W^2\Phi^4}
 +     \frac{f^{(1)}_{BW \Phi^4}}{\Lambda^4}   {\cal O}^{(1)}_{BW \Phi^4}
      + \frac{\Delta^{(8)}_{4F}}{\Lambda^4} {\cal O}_{4F}^{(8)} \;,
                       \label{eq:leff}
\end{eqnarray}
where $ {\cal O}_{4F}$ stands for the part of $\ocal_{2JW} $ that
contributes to the muon decay while $ {\cal O}_{4F}^{(8)}$ is the
corresponding dimension-eight operator. They have been defined so that
their contribution to the Higgs field vacuum expectation value in the
SM Lagrangian reads
\begin{equation}
  \left[2\langle \Phi^\dagger \Phi \rangle - \frac{1}{\sqrt{2} \hat{G}_F}
  \right]_{\rm fermionic}\equiv  \frac{\vh^4}{\Lambda^2} \Delta_{4F}
  + \frac{\hat{v}^6}{\Lambda^4} \Delta^{(8)}_{4F}  \;.
\label{eq:vt}
\end{equation}

The predictions for observables at order $1/\Lambda^4$ require
evaluating the SM contributions, the interference between the
$1/\Lambda^2$ amplitude (${\cal M}^{(6)}$) with the SM amplitude, the
square of the dimension-six amplitude, as well as the interference of
the dimension-eight amplitude ${\cal M}^{(8)}$ with the SM one, that
we represent as:
\begin{equation}
  | M_{\rm SM}|^2 + {\cal M}_{\rm SM}^\star {\cal M}^{(6)}  + | {\cal
    M}^{(6)}|^2 + {\cal M}_{\rm SM}^\star {\cal M}^{(8)}  \;.
\end{equation}
Notice that ${\cal M}^{(8)}$ includes dimension-eight vertices as well
as the contribution of the insertion of two dimension-six couplings in
the amplitude.  \medskip

\section{EWPO analysis}
\label{sec:ewpo}

Our EWPO analysis includes 14 observables of which 12 are $Z$
observables~\cite{ALEPH:2005ab}:
\begin{eqnarray*}
&&\Gamma_Z \;\;,\;\;
\sigma_{h}^{0} \;\;,\;\;
{\cal A}_{\ell}(\tau^{\rm pol}) \;\;,\;\;
R^0_\ell \;\;,\;\;
{\cal A}_{\ell}({\rm SLD}) \;\;,\;\;
A_{\rm FB}^{0,l} \;\;,\;\;
\\
 && R^0_c \;\;,\;\;
 R^0_b \;\;,\;\;
{\cal  A}_{c} \;\;,\;\;
 {\cal A}_{b} \;\;,\;\;
A_{\rm FB}^{0,c}\;\;,\;\;
\hbox{ and} \;\;
A_{\rm FB}^{0,b}  \hbox{ (SLD/LEP-I)}\;\;\; ,
\end{eqnarray*}
supplemented by two $W$ observables
\begin{equation*}
  M_W   \;\;,\;\; \Gamma_W \;\;
\end{equation*}
that are, respectively, its average $W$-boson mass taken
from~\cite{Olive:2016xmw}\footnote{In order to be conservative we did
  not take into account the recent CDF measurement of the $W$
  mass~\cite{CDF:2022hxs}.}, its width from
LEP2/Tevatron~\cite{ALEPH:2010aa}\footnote{We do not include the
  average leptonic $W$ branching ratio because it does not include any
  additional constraint for universal EFT.}.  The correlations among
these inputs~\cite{ALEPH:2005ab} are taken into consideration in the
analyses. The SM predictions and their uncertainties due to variations
of the SM parameters were extracted from~\cite{deBlas:2022hdk}.
\medskip

The statistical analysis of the EWPO data  is made by means of a binned
log-likelihood function defining a $\chi^2$ function which depends on
seven Wilson coefficients,
\begin{eqnarray}
  \chi^2_{\rm  EWPO}&\equiv&  \chi^2_{\rm EWPO}\left(f_{BW},f_{\Phi,1},
                             \Delta_{4F},
                             f^{(1)}_{BW\Phi^4},f^{(2)}_{D^2 \Phi^6}, \Delta_{4F}^{(8)},f^{(3)}_{W^2\Phi^4}\right) \; .
\label{eq:chiewpo}
\end{eqnarray}
In fact, EWPO cannot constrain the seven Wilson coefficients
independently.  This is so because, as described in the
Appendix~\ref{app:inputs}, the corrections to the $Z$ interaction to
fermions to order $\Lambda^{-4}$ can be expressed in terms of the
following three combinations of Wilson coefficients:
\begin{eqnarray}
  &&\widetilde \Delta_{4F} = \Delta_{4F}+ \frac{\vh^2}{\Lambda^2}
   \Delta_{4F}^{(8)}  \;, 
\nonumber
  \\
  && \widetilde{f}_{BW} =
  f_{BW} + \frac{\vh^2}{2\Lambda^2} f^{(1)}_{BW\Phi^4} \; ,
\label{eq:ewpotilde}
  \\
  && \widetilde{f}_{\Phi,1} = f_{\Phi,1} + \frac{\vh^2}{\Lambda^2}
     f^{(2)}_{D^2\Phi^6}  \;.
\nonumber
\end{eqnarray}
The corrections to the $W$ mass and coupling to fermions further involve
the addition of only one operator $\ocal^{(3)}_{W^2\Phi^4}$.  Using
these variables we incorporate in our calculation some higher order
terms in the $1/\Lambda$ expansion in the spirit of geometric
SMEFT~\cite{Helset:2020yio,Corbett:2021eux}. \medskip

These three coefficient combinations and $f^{(3)}_{W^2\phi^4}$ are
directly related to the contributions to the oblique $S$, $T$, $U$
parameters~\cite{Peskin:1990zt}, and $\delta G_F$ at linear order in
Wilson coefficients of operators up to dimension-eight:
\begin{equation}
  \alpha S= -\hat e^2 \frac{\vh^2}{\Lambda^2}\widetilde{f}_{BW}\; ,
  \;\;\;\;
  \alpha T= -\frac{\vh^2}{2\Lambda^2}\widetilde{f}_{\Phi,1}\; ,
  \;\;\;\;
  \alpha U=\hat e^2 \frac{\vh^4}{\Lambda^4 }f^{(3)}_{W^2\Phi^4} \;
  \;\;,\;\;
  \frac{\delta G_F}{\hat G_F}=\frac{\vh^2}{\Lambda^2}\widetilde{\Delta}_{4F}\; .
\end{equation}
It is interesting to notice that
there is a contribution to the oblique parameter $U$ at
dimension-eight. Thus, effectively the EWPO chi-squared function is:
\begin{eqnarray}
  \tilde\chi^2_{\rm  EWPO}&\equiv&  \tilde\chi^2_{\rm EWPO}(\tilde{f}_{BW},\tilde{f}_{\Phi,1},
  f^{(3)}_{W^2\Phi^4}
  , \widetilde{\Delta}_{4F}) \; .
\label{eq:chiewpo2}
\end{eqnarray}
%

\begin{figure}
\includegraphics[width=0.8\textwidth]{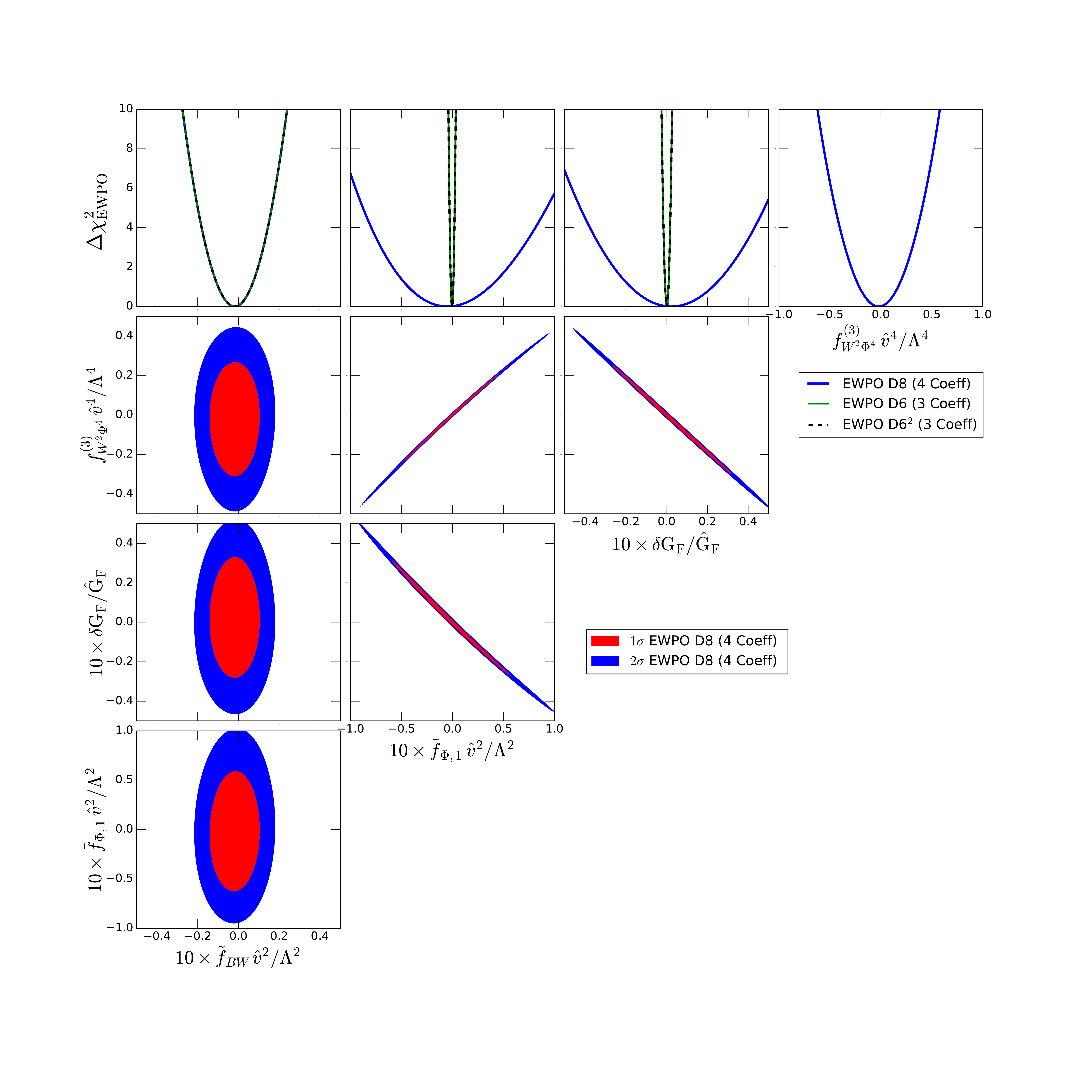}
\caption {One- and two-dimensional projections of
  $\Delta \tilde{\chi}^2_{\rm EWPO}$ for the 
  coefficients $\tilde{f}_{BW} \vh^2/\Lambda^2$,
  $\tilde{f}_{\Phi,1} \vh^2/\Lambda^2$,
  $\delta G_F/\hat{G}_F$, and
  $f^{(3)}_{W^2\Phi^4} \vh^4/\Lambda^4$, as indicated in each panel after
  marginalizing over the undisplayed parameters.}
 \label{fig:ewpd}
\end{figure}

Figure~\ref{fig:ewpd} shows the one- and two-dimensional projections
of $\Delta\tilde\chi^2_{\rm EWPO}$ as a function of the coefficients
$\widetilde{f}_{BW} \vh^2/\Lambda^2$,
$\widetilde{f}_{\Phi,1} \vh^2/\Lambda^2$, $\delta G_F/\hat{G}_F$, and
$f^{(3)}_{W^2\Phi^4} \vh^4/\Lambda^4$.  The panels in the top row contain
the one-dimension marginalized projection of
$\Delta\tilde\chi^2_{\rm EWPO}$, where the dashed line stands for the
$\ocal(1/\Lambda^2)$ analysis while the green solid one also contains
the dimension-six squared contribution. The full analysis that
includes the dimension-eight contribution is represented by the blue
line. As seen in the figure, the results at linear dimension-six and
dimension-six squared are identical, which is expected given the
precision of the data.\medskip

From Fig.~\ref{fig:ewpd} we also see that once the dimension-eight coefficient
$f^{(3)}_{W^2\Phi^4}$ is included the bounds on
$\widetilde{f}_{\Phi,1} \vh^2/\Lambda^2$ and $\delta G_F/\hat{G}_F$
weaken by about a factor 3--4.
The main reason is that when $f^{(3)}_{W^2\Phi^4}$ is also included in
the analysis cancellations can occur. In particular as can be seen in
Eqs.~\eqref{eq:dg1}--\eqref{eq:dmw} for
\begin{equation}
  \widetilde{f}_{\Phi,1}=-2\widetilde{\Delta}_{4F}=
  \frac{\hat e^2}{2\sh^2}\frac{\vh^2}{\Lambda^2 }f^{(3)}_{W^2\Phi^4}
  \;
  \label{eq:cancel}
\end{equation}
the linear contributions from $\widetilde{f}_{\Phi,1}$ ({\em i.e.}
$T$), $\widetilde \Delta_{4F}$ ($\delta G_F/\hat{G}_F$) and
$f^{(3)}_{W^2\Phi^4}$ ($U$), cancel both in the Z
observables and in $M_W$. Therefore, along this direction in the
parameter space, the bounds on these three quantities dominantly come
from the contribution of $\Gamma_W$ in Eq.~\eqref{eq:dgwn}, but this
observable is less precisely determined. Hence the 
strong correlations we
 observe in the corresponding two-dimensional allowed regions in
Fig.~\ref{fig:ewpd}. Nevertheless, the limits are
still quite stringent; see Table~\ref{tab:ranges_ewpo}.
\medskip

\begin{table} [ht]
  \centering
\begin{tabular}{|c|l|l|}
  \hline
       & \multicolumn{2}{c|}{EWPO 95\% CL allowed range} \\\hline
    Coupling & dimension 6  &  dimension 8  \\\hline
    $\frac{\vh^2}{\Lambda^2}\, \tilde f_{BW}$   &
    $[-0.018,0.044]$ &$[-0.018,0.044]$ \\[+0.1cm]
    $\frac{\vh^2}{\Lambda^2}\, \tilde f_{\Phi,1}$   &
    $[-0.0028,0.0018]$ &$[-0.080,0.081]$ \\[+0.1cm]
    $\frac{\delta G_F}{\hat{G}_F}$   &
    $[-0.0016,0.0017]$ &$[-0.038,0.044]$ \\[+0.1cm]
    $\frac{\vh^4}{\Lambda^4}\, f^{(3)}_{W^2\Phi^4}$   &
    \qquad --- &$[-0.40,0.36]$\\\hline 
\end{tabular}
\caption{95\% CL allowed ranges for the effective couplings entering
  in the EWPO with the analysis done including only the dimension-six
  contributions (left column) and also the dimension-eight
  contributions (right column).}
\label{tab:ranges_ewpo}
\end{table}

\section{Diboson analysis}
\label{sec:results}

The electroweak production of $WZ$, $WW$ and $W\gamma$ pairs, as well
as the vector boson fusion production of $Z$'s ($Zjj$), collectively
denoted by EWDB, allow us to study the triple couplings of electroweak
gauge bosons. In this work we consider the EWDB data shown in
Table~\ref{tab:tgv-data} which comprise a total of 73 data points. \medskip

\begin{table} [ht]
\begin{tabular}{|@{\hskip 0.5cm}c|l|l|c|l|l|}
\hline
& Channel ($a$) & Distribution & \# bins   &\hspace*{0.2cm} Data set & \hspace*{0.2cm}Int Lum  \\ [0mm]
  \hline
  & $WZ \to \ell^+ \ell^- \ell^{\prime\pm}$ & {$M(WZ)$} & 7& CMS 13
       TeV,  &  137.2 fb$^{-1}$~\cite{CMS:2021lix} \\[0mm]
  \multirow{7}{*}
{\begin{rotate}{90}   EWDB data\end{rotate}}
&$WW \to \ell^+\ell^{(\prime)-}+ 0/1 j$   &$M(\ell^+\ell^{(\prime)-})$
                             &11 & CMS 13 TeV, & 35.9 fb$^{-1}$~\cite{CMS:2020mxy} \\[0mm]
&  $W\gamma \to \ell \nu \gamma$ & $\frac{d^2\sigma}{dp_Td\phi}$ & 12
       & CMS 13 TeV, & 137.1 fb$^{-1}$~\cite{CMS:2021rym}
       \\[0mm]
&  $WW\rightarrow e^\pm \mu^\mp+\sla{E}_T\; (0j)$
&  $m_T$ & 17 (15) &
ATLAS 13 TeV, &36.1 fb$^{-1}$~\cite{Aaboud:2017gsl} \\[0mm]
& $WZ\rightarrow \ell^+\ell^{-}\ell^{(\prime)\pm}$
&  $m_{T}^{WZ}$ & 6
& ATLAS 13 TeV, &36.1 fb$^{-1}$~\cite{ATLAS:2018ogj} \\[0mm]
& $ Z jj \to \ell^+\ell^- jj$ & $\frac{d\sigma}{d\phi}$ & 12 & ATLAS 13 TeV,
& 139 fb$^{-1}$~\cite{ATLAS:2020nzk} \\[0mm]
& $WW\rightarrow \ell^+\ell^{(\prime) -}+\sla{E}_T\; (1j)$ &
$\frac{d\sigma}{dm_{\ell^+\ell^-}}$ & 10 & ATLAS 13 TeV,
 & 139 fb$^{-1}$~\cite{ATLAS:2021jgw} \\[0mm]
  \hline
\end{tabular}
\caption{EWDB data from LHC used in the analyses.  For
  the $W^+W^-$ results from ATLAS run 2~\cite{Aaboud:2017gsl} we
  combine the data from the last three bins into one to ensure
  gaussianity.}
\label{tab:tgv-data}
\end{table}

The theoretical predictions needed for the EWDB data are obtained by
simulating at leading order the $W^+W^-$, $W^\pm Z$, $W^\pm \gamma$,
and $Zjj$ channels that receive contributions from TGC.  To this end,
we use \textsc{MadGraph5\_aMC@NLO}~\cite{Frederix:2018nkq} with the
UFO files for our effective Lagrangian generated with
\textsc{FeynRules}~\cite{Christensen:2008py, Alloul:2013bka}.  We
employ \textsc{PYTHIA8}~\cite{Sjostrand:2007gs} to perform the parton
shower and hadronization, while the fast detector simulation is
carried out with \textsc{Delphes}~\cite{deFavereau:2013fsa}.  Jet
analyses are performed using
\textsc{FASTJET}~\cite{Cacciari:2011ma}.\medskip

The results of the analysis can be qualitatively understood in 
terms of the effective
$\gamma W^+ W^-$ and $Z W^+W^-$ TGC introduced in Ref.~\cite{Hagiwara:1986vm} 
\begin{eqnarray}
{\cal L}_{WWV} =&& 
 -i g_{WWV} \Big\{ 
g_1^V \Big( W^+_{\mu\nu} W^{- \, \mu} V^{\nu} 
  - W^+_{\mu} V_{\nu} W^{- \, \mu\nu} \Big)\nonumber\\ 
 &&+ \kappa_V W_\mu^+ W_\nu^- V^{\mu\nu}
+ \frac{\lambda_V}{\hat M_W^2} W^+_{\mu\nu} W^{- \, \nu\rho} V_\rho^{\; \mu}
 \Big\}
\;\;,
\label{eq:tgc}
\end{eqnarray}
where $g_{WW\gamma} = \eh$, $g_{WWZ} = \eh \ch / \sh$, and $\hat{M}_W = \eh \vh/2\sh$. In the SM
$g_1^\gamma=g_1^Z = \kappa_\gamma = \kappa_Z =1$ and
$\lambda_Z=\lambda_\gamma=0$. After including the direct contribution from  the
dimension-six and dimension-eight operators, electromagnetic gauge invariance still enforces $g_1^\gamma=1$, while the other effective TGC couplings
read:
\begin{eqnarray}
 &&\Delta  g_1^Z = \frac{\eh^2}{\sh^2 \ch^2} \left[ \frac{1}{8} \frac{\vh^2}{\Lambda^2}
\left ( f_W +   \frac{\vh^2}{2\Lambda^2} f^{(1)}_{W\Phi^4 D^2} \right
  ) \right] \;,
  \nonumber
  \\
  && \Delta \kappa_\gamma = \frac{\eh^2}{\sh^2} \left[ \frac{1}{8}
     \frac{\vh^2}{\Lambda^2} \left (  f_W +   \frac{\vh^2}{2\Lambda^2}
     f^{(1)}_{W\Phi^4 D^2} +  f_B + \frac{\vh^2}{2\Lambda^2} f^{(1)}_{B\Phi^4 D^2}
     \right)     \right] \;,
     \nonumber
  \\
  && \Delta \kappa_Z = \frac{\eh^2}{\sh^2} \left [ \frac{1}{8}  \frac{\vh^2}{\Lambda^2}  \left(
      f_W +   \frac{\vh^2}{2\Lambda^2}
     f^{(1)}_{W\Phi^4 D^2} \right) - \frac{\sh^2}{8\ch^2}
     \frac{\vh^2}{\Lambda^2} \left(  f_B + \frac{\vh^2}{2\Lambda^2} f^{(1)}_{B\Phi^4 D^2}
     \right)
     \right] \;,
     \label{eq:hagi}
  \\
  && \lambda_\gamma = \frac{3 \eh^2}{2\sh^2}
     \frac{\hat M_W^2}{\Lambda^2} \left[
     f_{WWW} + \frac{\vh^2}{2\Lambda^2} f^{(1)}_{W^3\Phi^2}
     \right]
     - \frac{\hat M_W^4} {2\Lambda^4} f^{(1)}_{W^2B\Phi^2} \;,
     \nonumber
     \\
    && \lambda_Z = \frac{3 \eh^2}{2\sh^2}
     \frac{\hat M_W^2}{\Lambda^2} \left[
     f_{WWW} + \frac{\vh^2}{2\Lambda^2} f^{(1)}_{W^3\Phi^2}
     \right]
      + \frac{\hat M_W^4} {2\Lambda^4} \frac{\sh^2}{\ch^2}
       f^{(1)}_{W^2B\Phi^2} \;.
     \nonumber
\end{eqnarray}

The complete analysis of diboson production at fixed order
$1/\Lambda^4$ depends on not only the direct SMEFT contributions to TGC
in \eqref{eq:hagi}, but also on the indirect contributions from
${\cal O}_{BW}$, ${\cal O}_{\Phi,1}$, ${\cal O}_{4F}$,
${\cal O}^{(1)}_{BW \Phi^4}$, ${\cal O}^{(2)}_{D^2 \Phi^6 }$, and
${\cal O}_{4F}^{(8)}$, through renormalization of the  SM gauge
couplings to fermions and TGC. In appendix~\ref{app:tgc} we list the
complete expressions and show that, in fact, the indirect effects
involve the same there combinations \eqref{eq:ewpotilde} and are
therefore bounded by the EWPO. In light of the constraints derived in
the previous section, in what follows we will neglect the effect of
those operators in the EWDB data analysis.\medskip

For the direct effects, Eq.~\eqref{eq:hagi} explicitly shows that the
contributions of the dimension-eight operators
$ {\cal O}^{(1)}_{W\Phi^4 D^2}$, $ {\cal O}^{(1)}_{B\Phi^4 D^2}$ and
$ {\cal O}^{(1)}_{W^3\Phi^2}$ to the TGC couplings have the same
structure of the contributions from the dimension-six operators $\ow$,
$\ob$ and $\owww$, respectively.  Conversely,
${\cal O}^{(1)}_{W^2 B \Phi^2}$ contributes a purely
$\ocal (1/\Lambda^4)$ to $\lambda_\gamma \ne \lambda_Z$.\medskip

Following an approach equivalent to that employed for the
analysis of EWPO we can define three effective coefficients,
\begin{eqnarray}
  &&\tfw =  f_W + \frac{\vh^2}{2
     \Lambda^2} f^{(1)}_{W\Phi^4 D^2} \;,
\nonumber
  \\
  &&\tfb = f_B +
     \frac{\vh^2}{2\Lambda^2} f^{(1)}_{B\Phi^4 D^2} \;,
     \label{eq:tilde}
  \\
  &&\tf_{WWW} = f_{WWW} +
     \frac{\vh^2}{2} f^{(1)}_{W^3\Phi^2} \;.
\nonumber  
\end{eqnarray}
which, together with $f^{(1)}_{W^2 B \Phi^2}$, effectively  parametrize
the relevant contributions to  the EWDB analysis. \medskip

Following this approach we perform the statistical analysis of the EWDB data
using a binned chi-squared function defined in terms of these
effective coefficients
\begin{equation}
\tilde\chi^2_{\rm EWDB}\left (\tfw, \tfb, \tf_{WWW}, f^{(1)}_{W^2 B \Phi^2}\right ) \;.
\end{equation}

\begin{center}
\begin{figure}
\includegraphics[width=0.8\textwidth]{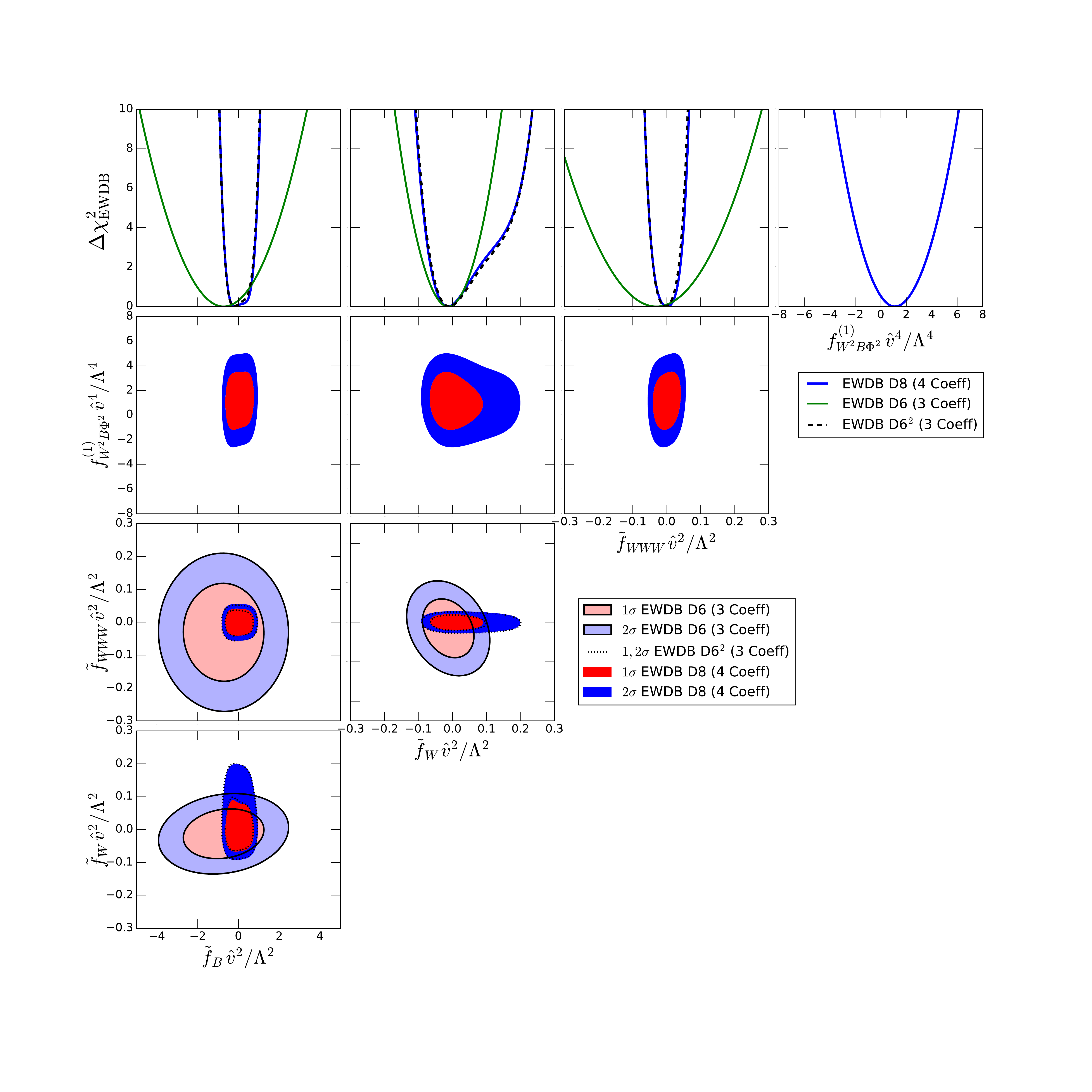}
\caption{
One- and two-dimensional projections of
$\Delta\tilde\chi^2_{\rm EWDB}$
  for the effective coefficients $\tfw \vh^2/\Lambda^2$,
  $\tfb \vh^2/\Lambda^2$, $\tf_{WWW} \vh^2/\Lambda^2$, and
    $f^{(1)}_{W^2 B \Phi^2}
    \vh^4/\Lambda^4$ as indicated in each panel
  after marginalizing over the undisplayed parameters.  }
 \label{fig:tgc-tilde-2d}
\end{figure}
\end{center}

Fig.~\ref{fig:tgc-tilde-2d} depicts the one- and two-dimensional
marginalized 68\% and 95\% CL allowed regions for
$\tfw \vh^2/\Lambda^2$, $\tfb \vh^2/\Lambda^2$,
$\tf_{WWW} \vh^2/\Lambda^2$, and
$f^{(1)}_{W^2 B \Phi^2} \vh^4/\Lambda^4$ after marginalizing over the
remaining fit parameters. The light pink (blue) regions in these
panels correspond to the 68\% (95\%) CL allowed regions of the
$\ocal( 1/\Lambda^2)$ analysis; see the three lower panels.  This
analysis yields the marginalized 95\% CL, allowed intervals for the
Wilson coefficients of the three relevant dimension-six operators
displayed in the left column of Table~\ref{tab:ranges_ewdbt}.
\medskip

\begin{table} [ht]
  \centering
\begin{tabular}{|c|l|l|l|}
  \hline
       & \multicolumn{3}{c|}{EWDB 95\% CL allowed range} \\\hline
    Coupling & dimension 6  & (dimension 6)$^2$& dimension 8  \\\hline
    $\frac{\vh^2}{\Lambda^2}\, \tilde f_{B}$   &
    $[-3.3,1.8]$ &
    $[-0.75,0.83]$ 
    &$[-0.73,0.86]$ \\[+0.1cm]
    $\frac{\vh^2}{\Lambda^2}\, \tilde f_{W}$   &
    $[-0.11,0.085]$
    &$[-0.079,0.16]$
    &$[-0.080,0.16]$
    \\[+0.1cm]
        $\frac{\vh^2}{\Lambda^2}\, \tilde{f}_{WWW}$   &
    $[-0.22,0.16]$
    &$[-0.049,0.045]$ 
    &$[-0.048,0.049]$
    \\[+0.1cm]
    $\frac{\vh^4}{\Lambda^4}\, f^{(1)}_{W^2B\Phi^2}$   &
    \qquad --- &\qquad --- &
    $[-1.9,4.2]$\\\hline 
\end{tabular}
\caption{95\% CL allowed ranges for the effective couplings entering
  in the EWDB analysis including only up to the dimension-six contributions
  (left column), up to the dimension-six squared contributions (central
  column) and including also the dimension-eight contributions (right
  column).}
\label{tab:ranges_ewdbt}
\end{table}

The dark red (blue) shaded regions in Fig.~\ref{fig:tgc-tilde-2d}
represent the two-dimensional allowed regions at 68\% (95\%) C.L.
including also the dimension-six squared and the dimension-eight
contributions. The corresponding one dimensional projections are given
in the blue lines in the upper panels.  For the sake of comparison we
also show  the corresponding results including only the dimension-six
squared contributions.  These are the black dashed lines in the
one-dimensional projections in the upper panels and the dotted line
contours in the three lower panels.  From the figure we see that
including the $1/\Lambda^4$ effects lead to stronger bounds on the
effective coefficients $\tilde{f}_B$ and $\tilde{f}_{WWW}$ while the
bound for $\tilde{f}_W$ is slightly looser and shifted; see also the
central and right columns of Table~\ref{tab:ranges_ewdbt}. 
We traced the counter-intuitive behaviour of the bounds on
$\tilde{f}_W$ to the $WZ$ datasets. Removing $WZ$ production from the fit  
leads to stronger limits at the $\ocal(1/\Lambda^4)$ also for
$\tilde{f}_W$.
\medskip

The results in Fig.~\ref{fig:tgc-tilde-2d} also show that the
dimension-six squared terms are dominant over the dimension-eight
one. Or in other words, the inclusion of the relevant dimension-eight
operator in this analysis, ${\cal O}^{(1)}_{W^2 B \Phi^2}$, has very
little impact on the results.  The physical reason for this can be
traced to the different dependence on the partonic center-of-mass
energy ($\hat{S}$) of the contribution to the relevant squared
amplitudes from dimension-six squared and dimension-eight terms. 
As it is well-known, the anomalous TGCs spoil the
cancellations that
  take place in the SM allowing the scattering amplitudes to grow with
  the partonic center-of-mass energy.  The fastest growing amplitudes
  are (for $\hat S\gg m_{W,Z}$): 
\begin{eqnarray}
  &&{\cal M}\left( d_- \bar{d}_+ \to W^+_0 W^ -_0 \right)=
  -i\frac{\eh^2}{24\sh^2\ch^2} \frac{\hat{S} }{\Lambda^2} \sin\theta
  \left[ 3 \ch^2 \,
    f_W     - \sh^2\,
    f_B + \frac{\vh^2}{2\Lambda^2}\left(
3 \ch^2 \, f^{(1)}_{W\Phi^4 D^2} 
- \sh^2\, f^{(1)}_{B\Phi^4 D^2} 
     \right) \right] \;,
     \nonumber
  \\
&&{\cal M}\left( d_+ \bar{d}_- \to W^+_\pm W^ -_\pm \right) =
      { i \frac{\eh^2}{48\sh^2\ch^2}   
        \frac{\hat{S}}{\Lambda^2}  \sin\theta \,
        \frac{\vh^2}{\Lambda^2}f^{(1)}_{W^2B\Phi^2}}\;,
   \nonumber
  \\
&& {\cal M} \left( d_- \bar{d}_+ \to W^+_\pm W^ -_\pm
    \right) =  -i\frac{3\eh^4}{8\sh^4} \frac{\hat{S}}{\Lambda^2} \sin\theta
    \left[f_{WWW} +\frac{\vh^2}{2\Lambda^2}\left(  f^{(1)}_{W^3\Phi^2}  
      + {\frac{\sh^2}{18\ch^2}} f^{(1)}_{W^2B\Phi^2}\right)\right]\;,
\nonumber
  \\
  && {\cal M} \left( d_+ \bar{d}_- \to W^+_0 W^-_0\right)=
     -i\frac{\eh^2}{12\ch^2} \frac{\hat{S}}{\Lambda^2} \sin\theta
    \left( f_B + \frac{\vh^2}{2\Lambda^2} f^{(1)}_{B\Phi^4 D^2}\right) \;,
     \nonumber 
  \\
  &&{\cal M} \left( u_-\bar{u}_+\to W^+_0W^-_0 \right) = i\frac{\eh^2}{24 \sh^2\ch^2}
  \frac{\hat{S}}{\Lambda^2} \sin\theta
  \left[ 3 \ch^2 \,
    f_W     + \sh^2\,
    f_B + \frac{\vh^2}{2\Lambda^2}\left(
3 \ch^2 \, f^{(1)}_{W\Phi^4 D^2} 
+ \sh^2\, f^{(1)}_{B\Phi^4 D^2} 
     \right) \right] \;,
     \nonumber
  \\
&& {\cal M} \left(
  u_+\bar{u}_-\to W^+W^-\right) =i \frac{\eh^2}{6\ch^2}
   \frac{\hat{S}}{\Lambda^2} \sin\theta
 \left( f_B + \frac{\vh^2}{2\Lambda^2} f^{(1)}_{B\Phi^4 D^2}\right) \;,
  \\
&& {\cal M} \left ( u_+ \bar{u}_- \to W^+_\pm W^-_\pm\right)= { -i
    \frac{\eh^4}{24\sh^2\ch^2}  
    \frac{\hat{S}}{\Lambda^2} \sin\theta\, \frac{\vh^2}{\Lambda^2}
                                             f^{(1)}_{W^2B\Phi^2} } \;,
\nonumber
  \\
  &&{\cal M} \left(   u_- \bar{u}_+ \to W^+_\pm W^-_\pm \right) = {
    i \frac{3\eh^4}{8\sh^4}
    \frac{\hat{S}}{\Lambda^2} \sin\theta \left[
       f_{WWW} +
       \frac{\vh^2}{2\Lambda^2} \left(  \,f^{(1)}_{W^3\Phi^2}  
      -  \frac{\sh^2}{18\ch^2} f^{(1)}_{W^2B\Phi^2}\right) 
\right]}\;,
     \nonumber
     \end{eqnarray}
     as well as
     \begin{eqnarray}
&&{\cal M} \left( d_-\bar{u}_+ \to Z_\pm  W^-_\pm\right) =
     i \frac{3 \ch \eh^4}{4\sqrt{2} \sh^4} \frac{\hat{S}}{\Lambda^2} \sin\theta\left[
 f_{WWW} +\frac{\vh^2}{2\Lambda^2} \left(f^{(1)}_{W^3\Phi^2}  
 +{ \frac{\sh^2}{6 \ch^2} 
    f^{(1)}_{W^2B\Phi^2} }\right)\right]\;,
\nonumber
  \\
  &&{\cal M}  \left( d_-\bar{u}_+ \to Z_0  W^-_0 \right) = i
     \frac{\eh^2}{4\sqrt{2}\sh^2} \frac{\hat{S}}{\Lambda^2} \sin\theta
\left(  f_W + \frac{\vh^2}{2\Lambda^2} f^{(1)}_{W\Phi^4 D^2} \right) \;,
  \\
  && {\cal M} \left ( d_-\bar{u}_+ \to \gamma_\pm  W^-_\pm\right) = i
     \frac{3\eh^4}{4   \sqrt{2} \sh^3} \frac{\hat{S}}{\Lambda^2} \sin\theta\left[
f_{WWW} +\frac{\vh^2}{2\Lambda^2} \left(f^{(1)}_{W^3\Phi^2}   -
{ \frac{1}{6}  f^{(1)}_{W^2B\Phi^2}} \right)
  \right]\;,
  \nonumber
\end{eqnarray}
where we indicated the particle polarization as a subscript and
we denoted by $\theta$ the polar scattering angle in the center of
mass system. 
Therefore, dimension-six squared contribution
to the amplitude squared grows as
$\hat{S}^2$, while  the
dimension-eight contribution -- which enters in the interference with
the SM amplitude -- grows
as $\hat{S}$. \medskip

Notice that, unlike EWPO, which correspond to squared amplitudes for
fixed center-of-mass energy (either $M_Z$ or $M_W$), EWDB data
correspond to squared amplitudes at different center-of-mass energies.
Thus, since at order $1/\Lambda^4$, the dimension-six squared and
the dimension-eight contributions exhibit different energy dependence,
the approximate analysis performed in terms of the effective couplings
\eqref{eq:tilde}, does not exhaust the potential of the data to constrain the
Wilson coefficients of all the operators involved.
It is then possible to  perform an analysis in terms of the
seven Wilson coefficients contributing to the amplitudes of the
EWDB data because in fact to order $1/\Lambda^4$
the $\chi^2$ function depends independently on them:
\begin{equation}
  \chi^2_{\rm EWDB}\left(f_W, f_B, f_{WWW}, f^{(1)}_{W^2 B \Phi^2},
  f^{(1)}_{W\Phi^4D^2}, f^{(1)}_{B\Phi^4D^2}, f^{(1)}_{W^3\Phi^2} \right)
  \;.
  \label{eq:ch2-68}
\end{equation}

\begin{figure}
\includegraphics[width=\textwidth]{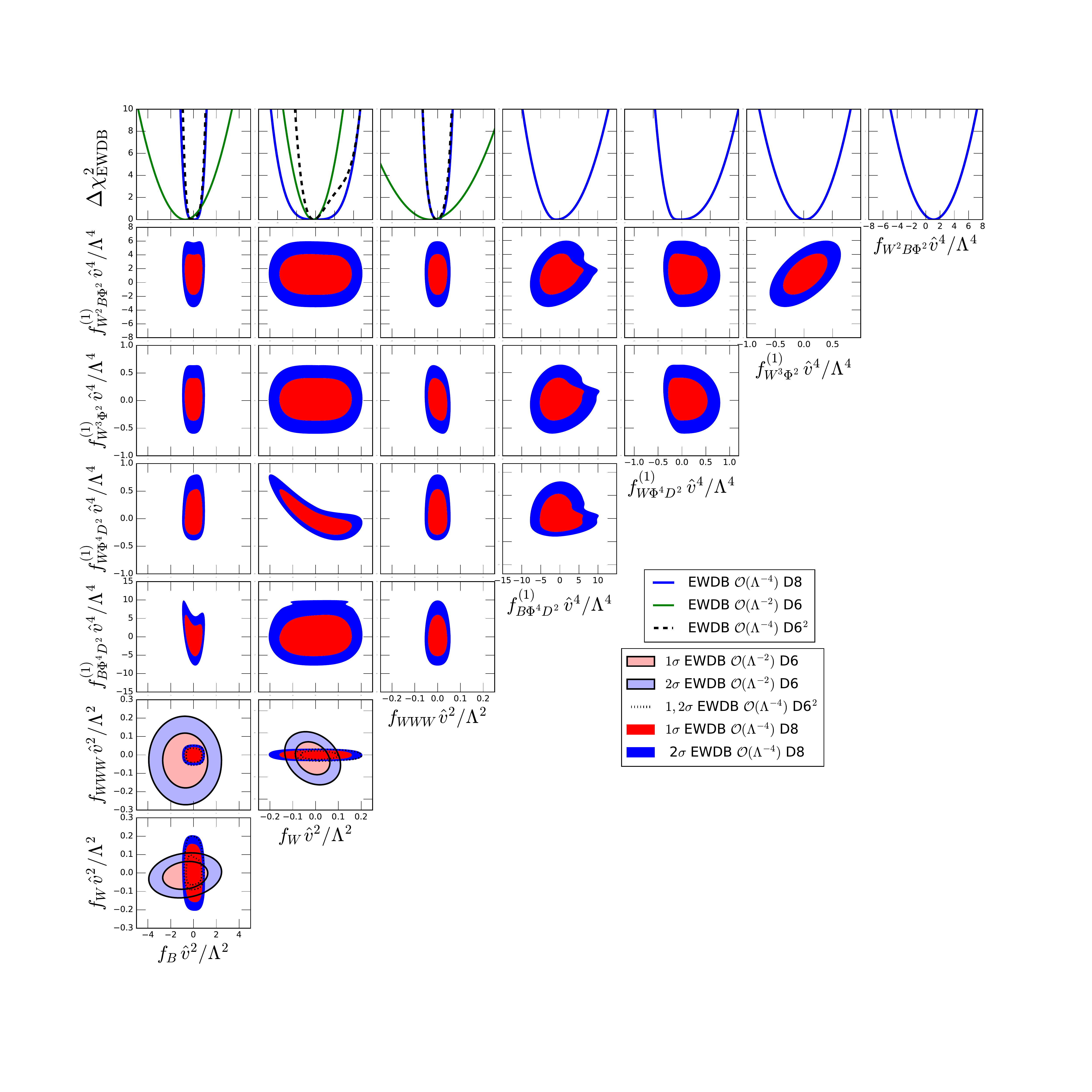}
\caption{One- and two-dimensional 68\% and 95\% CL projections of
  $\Delta\chi^2_{\rm EWDB}$ for $f_B \vh^2/\Lambda^2$,
  $f_W \vh^2/\Lambda^2$, $f_{WWW} \vh^2/\Lambda^2$,
  $f^{(1)}_{B\Phi^4D^2} \vh^2/\Lambda^4$,
  $f^{(1)}_{W\Phi^4D^2} \vh^2/\Lambda^4$,
  $f^{(1)}_{W^3\Phi^2} \vh^2/\Lambda^4$, and
  $f^{(1)}_{W^2B\Phi^2} \vh^2/\Lambda^4$ as indicated in the panels
  after marginalizing over the remaining fit parameters.  }
 \label{fig:tgc6-2d}
\end{figure}

We present in Fig.~\ref{fig:tgc6-2d} the one- and two-dimensional
marginalized 68\% and 95\% C.L. allowed regions for the seven Wilson
coefficients in Eq.~\eqref{eq:ch2-68} for several analyses differing
by the order in $1/\Lambda$ used in the calculations. We list the
corresponding 95\% CL allowed ranges in
Table~\ref{tab:limits-tgc6-lam4}.  Notice that the
$\ocal(1/\Lambda^2)$, [$\ocal(1/\Lambda^4)$ (dim-6)$^2$] analysis is
identical to the one described above as {\sl dimension-6} [{\sl
  (dimension-6)$^2$}] and leads to the limits on the Wilson
coefficients $f_W$, $f_B$ and $f_{WWW}$ given on left (central) column
in Table~\ref{tab:ranges_ewdbt} and the light shaded regions in
Fig.~\ref{fig:tgc-tilde-2d} (dotted contours) in the three lowest
panels.  We reproduce these regions and ranges in
Fig.~\ref{fig:tgc6-2d} and Table~\ref{tab:limits-tgc6-lam4} for
clarity and completeness.\medskip

From the figure we see that including the ${\cal O}(1/\Lambda^4)$
terms strengthens the constraints obtained at order $1/\Lambda^2$ for
the operators $\ob$ and $\owww$ while it weakens the bounds on $\ow$,
as expected from the results obtained in the approximate analysis in
Fig.~\ref{fig:tgc-tilde-2d}. \medskip

The comparison with the analysis performed including only
dimension-six squared terms in the evaluation of the $1/\Lambda^4$
contribution (see dashed lines) shows that the dimension-six Wilson
coefficient whose determination is most quantitatively affected by the
inclusion of the independent effects of the four dimension-eight
Wilson coefficients is $f_W$.  The reason for this is the
anti-correlation between $f_W$ and $f^{(1)}_{W\Phi^4D^2}$ that is
apparent in the second panel of the fourth row; see
Eq.~\eqref{eq:hagi}. In other words, the EWDB data analyzed provides a
weaker discrimination between the dimension-six and the
dimension-eight contribution to $\tilde f_W$.  On the contrary the
corresponding two-dimensional plots in Fig.~\ref{fig:tgc6-2d} show
that no large correlations are present between $f_B$ and
$f^{(1)}_{B\Phi^4D^2}$, nor between $f_{WWW}$ and $f_{W^3B \Phi^2}$.
The only other large correlation is observed between the
dimension-eight coefficients $f^{(1)}_{W^2B \Phi^2}$ and
$f_{W^3 \Phi^2}$ both contributing at the same order to
$\lambda_\gamma$ and $\lambda_Z$.  At the linear order on these
coefficients the stronger sensitivity comes from the $W\gamma$ channel
which bounds the combination $6 f_{W^3 \Phi^2}-f^{(1)}_{W^2B \Phi^2}$
(see Eq.~\eqref{eq:hagi}) leading to the positive correlation
observed. \medskip

\begin{table}[h!]
\centering
\begin{tabular}{|c|l|l|l|}\hline
Coefficient & \multicolumn{3}{c|}{EWPB 95\% CL allowed range}  \\\hline
& ${\cal O}(\Lambda^{-2})$ & ${\cal O}(\Lambda^{-4})$ (dim-6)$^2$
& ${\cal O}(\Lambda^{-4})$ (dim-6)$^2$ + dim-8\\\hline
    $\frac{\vh^2}{\Lambda^2}\, f_{B}$   &
    $[-3.3,1.8]$ &
    $[-0.75,0.83]$ 
   &\qquad $[-0.89, 0.89]$\\[+0.1cm]
$\frac{\vh^2}{\Lambda^2}\,f_{W}$   &
    $[-0.11,0.085]$
&$[-0.079,0.16]$
& \qquad $[-0.18, 0.18]$
    \\[+0.1cm]
    $\frac{\vh^2}{\Lambda^2}\,{f}_{WWW}$   &
    $[-0.22,0.16]$
    &     $[-0.049,0.045]$
    &\qquad$[-0.05,0.05]$\\[+0.1cm]
    $\frac{\vh^4}{\Lambda^4}\, f^{(1)}_{W^2B\Phi^2}$   &
    \qquad --- &\qquad \qquad --- 
& \qquad $[-2.6, 5.0]$\\[+0.1cm]
    $\frac{\vh^4}{\Lambda^4}\, f_{B\Phi^4D^2}$
&    \qquad --- &\qquad \qquad--- 
& \qquad $[-6.48, 7.8]$\\
    $\frac{\vh^4}{\Lambda^4}\, f_{W\Phi^4D^2}$
    &    \qquad --- &\qquad\qquad
    --- 
& \qquad $[-0.33, 0.66]$\\
$\frac{\vh^4}{\Lambda^4}\, f_{W^3\Phi^2}$
    &  \qquad --- &\qquad --- 
& \qquad $[-0.47, 0.51]$\\\hline
\end{tabular}
\caption{95\% C.L. allowed range
  for the Wilson coefficients present in the EWDB data analysis
  performed with predictions obtained at different orders in the
  $1/\Lambda^2$ expansion.}
  \label{tab:limits-tgc6-lam4}
\end{table}

\section{Summary and Conclusions}
\label{sec:disc}

We have studied the impact of $\ocal(1/\Lambda^4)$ corrections in the
EWPO and EWDB data analyses assuming a universal, C and P conserving
new physics scenario. The universality assumption reduces the number of
dimension-eight operators contributing to the processes making a
complete analysis possible. As described in Sec.~\ref{sec:model}, in
the HISZ basis for the dimension-six SMEFT the universal theories are
described by 11 bosonic operators and five fermionic operators, the
latter being generated by the application of the EOM in the reduction
of the basis.  At dimension eight there are ten potentially relevant
bosonic operators and one expects a fermionic operator generated by
the EOM.  Of those, we find that there are six (nine) dimension-six
(-eight) operators contributing to the EWPO and EWDB observables (see
Eq.~\eqref{eq:leff}). \medskip

The analysis of EWPO involves three dimension-six and four
dimension-eight operators whose Wilson coefficients cannot be
independently bound.  However, we find that it is possible to
eliminate the blind directions by redefining three effective
coefficients which are just a shift of the three Wilson coefficients
of the dimension-six operators corrected by their corresponding
dimension-eight siblings -- see Eq.~\eqref{eq:ewpotilde}-- and which
contain the sibling dimension-eight contribution to the universal
parameters S, T, and $\Delta G_F$.  In addition the analysis contains
a purely dimension-eight contribution to the universal parameter $U$.
The fit to EWPO performed in terms of these four parameters results in
strong constraints on $\tilde{f}_{BW}$, $\tilde{f}_{\Phi,1}$,
$\tilde{\Delta}_{4F}$, and $f^{(3)}_{W^2\Phi^ 4}$; see
Table~\ref{tab:ranges_ewpo}. \medskip

At $\ocal(1/\Lambda^4)$ EWDB analysis involves six (seven)
dimension-six (-eight) operators of which three (four) contribute
directly to the TGC while three (three) enter indirectly via the
finite renormalization of the SM parameters.  The indirect
contributions can be cast in terms of three effective couplings
bounded by the EWPO (see Appendix~\ref{app:tgc}) allowing us to
neglect them in the EWDB analysis.  \medskip

The direct contributions to the TGC can be  expressed  in
terms of three effective coefficients which are just a shift of the
three Wilson coefficients of the corresponding three dimension-six
operators corrected by their corresponding dimension-eight siblings
(i.e. rescaled by $\Phi^\dagger\Phi$); see Eq.~\eqref{eq:tilde}. In
addition there is a genuine dimension-eight contribution to the
difference between the $\lambda_\gamma$ and $\lambda_Z$ couplings. We
performed an effective analysis of the EWDB data in terms of these
four coefficients and showed that the bulk of $\ocal(1/\Lambda^4)$
impact on the analysis is due to the dimension-six squared
contribution $|{\cal M}^{(6)}|^2$; see Fig.~\ref{fig:tgc-tilde-2d} and
Table~\ref{tab:ranges_ewdbt}.  This is so because of the different
dependence on the partonic center-of-mass energy of dimension-six
squared terms which give a pure quadratic TGC contribution to the
amplitude squared, and the dimension-eight contribution which enters
in the interference with the SM amplitude.  \medskip

Profiting from the different energy dependence of the dimension-six
squared and the dimension-eight contributions it is possible to
perform an analysis of the EWDB data which allows us to constrain the
seven Wilson coefficients independently. The result of this analysis
is presented in Fig.~\ref{fig:tgc6-2d} and
Table~\ref{tab:limits-tgc6-lam4}.  The results show that the bounds on
the three Wilson coefficients of the dimension-six operators are only
slightly looser than in the effective four-parameter analysis, while
the bounds on the four Wilson coefficients of the dimension-eight
operators are all of similar order and all substantially weaker than
those on their dimension-six siblings. \medskip

In summary, we have shown that, for the universal scenario, the analysis of the EWPO and
the EWDB data allows us to constrain the full parameter space of
operators up to ${\cal O}(\Lambda^{-4})$. Within the present precision
of EWDB data and with our choice of basis, it is still consistent
to perform the analysis sequentially: first obtain the constraints
on the relevant Wilson coefficients using the EWPO and then apply those
bounds to reduce the number of Wilson coefficients which are relevant for the
the diboson analysis. That said, the LHC continues to accumulate data on
EWDB production, and consequently, we anticipate stronger bounds on the
TGC couplings in the future. At some point, it will be necessary
to perform a combined analysis of EWPO+EWDB data taking into account
the indirect contributions due to the finite renormalization to the TGC
in analogy with study of TGC and possible anomalous fermionic
couplings~\cite{Alves:2018nof}. \medskip

\acknowledgments

OJPE thanks the hospitality of the Departament de Fisica Quantica i
Astrofisica, Universitat de Barcelona, where part of this work was
carried out. OJPE is partially supported by CNPq grant number
305762/2019-2 and FAPESP grants 2019/04837-9 and 2022/05332-0.
M.M. is supported by FAPESP grant 2021/08669-3 while P.R. acknowledges
support by FAPESP grants 2020/10004-7 and 2021/12305-7.
This project is funded by USA-NSF grant PHY-1915093.  It has also
received support from the European Union's Horizon 2020
research and innovation program under the Marie Sk\l odowska-Curie
grant agreement No 860881-HIDDeN, and Horizon Europe research and
innovation programme under the Marie Sk\l odowska-Curie Staff Exchange
grant agreement No 101086085 -- ASYMMETRY''.
It also receives support from grants PID2019-105614GB-C21, and
``Unit of Excellence Maria de
Maeztu 2020-2023'' award to the ICC-UB CEX2019-000918-M, funded by
MCIN/AEI/10.13039/501100011033, and from
grant 2021-SGR-249 (Generalitat de Catalunya).
This manuscript has been authored by Fermi Research Alliance, LLC under
Contract No. DE-AC02-07CH11359 with the U.S. Department of Energy,
Office of Science, Office of High Energy Physics.

\newpage
\appendix

\section{Corrections to the $Z$ and $W$ couplings}
\label{app:inputs}

We parametrize the $Z$ coupling to fermion ($f$) pairs as 
\begin{equation}
  \frac{\eh}{\sh\ch} ~\left (\hat{g}^f\, (1+\Delta g_1)+
  Q^f\,\Delta g_2\right)
\end{equation}
where $\hat g^f=T_3^f-\sh^2 Q^f$, 
$T_3^f$ is the fermion third component of isospin and $Q^f$ is its
charge. After the renormalization of the SM parameters, we obtain at
order $1/\Lambda^4$
\begin{eqnarray}
  \Delta g_1
  = &&- \frac{1}{4} \frac{\vh^2}{\Lambda^2}
  \left[ 2 \left(\Delta_{4F} + \frac{\vh^2}{\Lambda^2} \Delta^{(8)}_{4F}\right)
    + f_{\Phi,1} +\frac{\vh^2}{\Lambda^2} 
    f^{(2)}_{D^2\Phi^6}
    \right]
  \nonumber
  \\ &&
  - \frac{1}{32} \frac{\vh^4}{\Lambda^4} \left[
  - 12 (\Delta_{4F})^2 + 4 \Delta_{4F}  f_{\Phi,1} - 3  (f_{\Phi,1})^2
  \right] \nonumber \\
    \simeq &&- \frac{1}{4} \frac{\vh^2}{\Lambda^2}
  \left[ 2 \tilde \Delta_{4F} 
    + \tilde f_{\Phi,1} \right]
  - \frac{1}{32} \frac{\vh^4}{\Lambda^4} \left[
  - 12 (\tilde \Delta_{4F})^2 + 4 \tilde \Delta_{4F}  \tilde f_{\Phi,1} - 3  (\tilde f_{\Phi,1})^2
        \right]   \;.
\label{eq:dg1}
\end{eqnarray}
In the last line we have used that to order $1/\Lambda^4$ the
corrections linear in the Wilson coefficients depend on
the four combinations in Eq.~(\ref{eq:ewpotilde}) and
therefore neglecting terms of $\ocal(1/\Lambda^6)$ we can rewrite
$\Delta g_1$ in terms of those combinations.
In the same way we find:
\begin{eqnarray}
  \Delta g_2=
  &&\frac{\vh^2}{\Lambda^2} \frac{1}{2\ch_2} \Big[
  -\sh^2\ch^2 \left(2 \tilde \Delta_{4F} 
  + \tilde f_{\Phi,1}\right)
  +
  \frac{\eh^2}{2} 
  \tilde f_{BW} \Big]
\nonumber
  \\ &&
    +\frac{\hat{v}^4}{\Lambda^4}\frac{1}{8\hat{c}_2^3}\Bigg\{\frac{\hat{s}_2^2}{4} \Big[ (1+3\hat{c}_4)\left((\tilde{\Delta}_{4F})^2+\frac14 (\tilde{f}_{\phi,1})^2\right)-(3+\hat{c}_4)\tilde{\Delta}_{4F}\tilde{f}_{\phi,1}\Big]\nonumber \\
    &&-\frac{\eh^2}{2}\left(\hat{c}_4\tilde{f}_{BW}\tilde{f}_{\phi,1}-2\tilde{\Delta}_{4F}\tilde{f}_{BW}+\eh^2(\tilde{f}_{BW})^2\right) \Bigg\}
\label{eq:dg2n}
\end{eqnarray}
with $\hat{c}_{n}=\cos (n\hat{\theta})$ and $\hat{s}_{n}=\sin (n\hat{\theta})$.

 \medskip

As for the $W$ observables 
\begin{eqnarray}
  \frac{\Delta  M_W}{\hat M_W} =
        && \frac{1}{4\ch_2} \frac{\vh^2}{\Lambda^2} \left [
      \eh^2 \tilde f_{BW} 
      - 2 \sh^2\tilde \Delta_{4F}        -\ch^2\tilde f_{\Phi,1}
              \right]+\frac{\eh^2}{8\sh^2} \frac{\vh^4}{\Lambda^4}  f^{(3)}_{W^2\Phi^4} \;
           \nonumber
   \\
        && +\frac{1}{8\ch^3_2}\frac{\vh^4}{\Lambda^4}
           \Big[ - \sh^4(2 + 3 \ch_2) (\tilde \Delta_{4F})^2
           + \frac{1}{4} \ch^4 (-2+5\ch_2) (\tilde f_{\Phi,1})^2
           -\frac{1}{16} \eh^4 \frac{(7-6\ch_2 + 3 \ch_4)}{\sh^2}  (\tilde f_{BW})^2
\nonumber
  \\
        && -\frac{\hat{c}^2}{4}  (9 - 6 \ch_2 + 5 \ch_4) \tilde \Delta_{4F} \tilde f_{\Phi,1}
          + \frac{1}{4} \eh^2 (7 - 2\ch_2 + 3\ch_4) \tilde \Delta_{4F}  \tilde f_{BW}
           - \frac{1}{2} \eh^2 \ch^2 ( -2+3\ch_2) \tilde f_{\Phi,1} \tilde f_{BW}\Big]           
\label{eq:dmw}
\end{eqnarray}
where $\hat M_W=\frac{\displaystyle \eh \vh}{\displaystyle 2\sh}$.
And we parametrize the $W$ coupling to left-handed fermions as
\begin{equation}
  \frac{\eh}{\sh} ~(1+\Delta g_W)
\end{equation}
where
\begin{eqnarray}
 \Delta g_W=&&
\frac{1}{4\ch_2} \frac{\vh^2}{\Lambda^2} \left [
      \eh^2 \tilde{f}_{BW}
        -2 \ch^2 \tilde\Delta_{4F}
        -\ch^2\tilde{f}_{\Phi,1}        \right]
\nonumber \\
&&+\frac{1}{8\ch_2^3} \frac{\vh^4}{\Lambda^4} \Big[
   \eh^2 \frac{\ch_2^3}{\sh^2} f^{(3)}_{W^2\Phi^4}
   + \ch^4 (-2+5 \ch_2) (\tilde{\Delta}_{4F})^2 - \frac{1}{16} \frac{(7 -6
   \ch_2 + 3 \ch_4)}{\sh^2}\eh^4 (\tilde{f}_{BW})^2 \nonumber
  \\
  && +\frac{1}{4} \ch^4 (-2+5 \ch_2) (\tilde{f}_{\Phi,1})^2 - \frac{1}{4}
     \ch^2 (7-6 \ch_2+3 \ch_4) \tilde{\Delta}_{4F} \tilde{f}_{\Phi,1} + \frac{1}{4}
     \eh^2 (5 - 2 \ch_2 + \ch_4) \tilde{\Delta}_{4F} \tilde{f}_{BW}
     \nonumber
  \\
  && - \frac{1}{2} \eh^2 \ch^2 (-2+ 3 \ch_2) \tilde{f}_{\Phi,1} \tilde{f}_{BW}\Big]
\label{eq:dgwn}
\end{eqnarray}

\section{Corrections to TGC}
\label{app:tgc}

The renormalization of the SM parameters give rise to indirect
contributions to TGC in addition to the direct contributions from the
dimension-six and -eight operators to the TGC. Using the parametrization for the
$\gamma W^+ W^-$ and $Z W^+W^-$ TGC given in Eq.~(\ref{eq:tgc}), we find 
that up to order $1/\Lambda^4$ (and neglecting terms of $\ocal(1/\Lambda^6)$)
the coupling to
$W^+_{\mu\nu} W^{- \, \mu} Z^{\nu}$ reads
\begin{eqnarray}
  g_1^Z = &&1 + \frac{1}{2} \frac{\vh^2}{\Lambda^2} 
  \left[  \frac{\eh^2}{4\sh^2\ch^2} \left(f_W +
         \frac{\vh^2}{2\Lambda^2} f^{(1)}_{W\Phi^4D^2} \right)
             - \frac{1}{\ch_2}\tilde\Delta_{4F}
  + \frac{1}{2} \frac{\eh^2}{\ch^2 \ch_2} \tilde f_{BW}
  - \frac{1}{2\ch_2} \tilde f_{\Phi,1} 
  \right]
  \nonumber
  \\
          && +\frac{1}{16\ch^3_2} \frac{\vh^4}{\Lambda^4} \Big[
             (1 + 2 \ch_2 + 3 \ch_4) \left( (\tilde \Delta_{4F})^2 +
             \frac{1}{4} (\tilde f_{\Phi,1})^2\right) - \frac{\eh^4}{\ch^2}
             (\tilde f_{BW})^2
             \nonumber
  \\
          && + 2 \frac{\eh^2}{\ch^2} \tilde \Delta_{4F} \tilde f_{BW}
             - (3 - 2 \ch_2 + \ch_4)\tilde \Delta_{4F}\tilde f_{\Phi,1}
             -
     \eh^2\frac{\ch_4}{\ch^2} \tilde f_{BW} \tilde f_{\Phi,1} \Big]
             \nonumber
  \\
  && -\frac{\eh^2}{4 \sh \ch \sh_4} \frac{\vh^4}{\Lambda^4}  \left(
     \tilde \Delta_{4F} - \eh^2\tilde f_{BW} + \frac{1}{2}  (1 +
     2 \ch_2) \tilde f_{\Phi,1}
     \right) f_W \;,
\end{eqnarray}
The couplings to $W_\mu^+ W_\nu^- V^{\mu\nu}$ are respectively 
\begin{eqnarray}
  \kappa_\gamma = &&1 + \frac{1}{8} \frac{\eh^2}{\sh^2}
                     \frac{\vh^2}{\Lambda^2} \left [ \left(f_B+
                     \frac{\vh^2}{2\Lambda^2} f^{(1)}_{B\Phi^4D^2}\right) +\left( f_W +
                     \frac{\vh^2}{2\Lambda^2} f^{(1)}_{W\Phi^4D^2}\right)
- 2
                     \tilde f_{BW} \right]
  \\
  && -\frac{\eh^2}{32} \frac{\vh^4}{\Lambda^4} \frac{1}{\sh^2 
     \ch_2}  \Big (
     2 (1-\ch_2) \tilde{\Delta}_{4F} - 2 \eh^2 \tilde{f}_{BW} + (1+\ch_2) \tilde{f}_{\Phi,1}
     \Big)
     (f_B + f_W- 2 \tilde f_{BW})
     \nonumber
  \\
                  &&
                     + \frac{\eh^2}{4 \sh^2}\frac{\vh^4}{\Lambda^4} f^{(3)}_{W^2\Phi^4}\,
         \nonumber 
\end{eqnarray}
and 
\begin{eqnarray}
  \kappa_Z = &&1 + \frac{1}{8} \frac{\eh^2}{\sh^2}
                \frac{\vh^2}{\Lambda^2} \left[
                \left( f_W + \frac{\vh^2}{2\Lambda^2} f^{(1)}_{W\Phi^4D^2}\right) 
                -\frac{\sh^2}{\ch^2} \left( f_B + \frac{\vh^2}{2\Lambda^2} f^{(1)}_{B\Phi^4D^2}\right) 
       + \frac{4\sh^2}{\ch_2} \tilde f_{BW}
                - \frac{4\sh^2}{\eh^2\ch_2} \tilde\Delta_{4F} 
   - \frac{2\sh^2}{\eh^2\ch_2} \tilde f_{\Phi,1}
   \right]
   \nonumber
   \\
    && +\frac{1}{16 \sh^2 \ch_2^3} \frac{\vh^4}{\Lambda^4}\left[
       \sh^2 (1+2\ch_2 + 3\ch_4) \left ((\tilde\Delta_{4F})^2 + \frac{1}{4}
       (\tilde f_{\Phi,1})^2 \right) - \eh^4 (2 -2 \ch_2 + \ch_4)
       (\tilde f_{BW})^2 
       \right.
       \nonumber
  \\
             && + \sh^2 (3 - 2 \ch_2 + \ch_4)\left( 2\eh^2\tilde
                \Delta_{4F} \tilde f_{BW}-\tilde{\Delta}_{4F}\tilde{f}_{\phi,1}\right)
                - \eh^2 \sh^2 (-1 + 2 \ch_2 + \ch_4) \tilde f_{BW} 
                \tilde f_{\Phi,1}
  \\
  && + f_W \left ( \eh^2 \ch_2^2 (-2 + \ch_2)
     \tilde \Delta_{4F}  +  \eh^2\frac{\ch_2^2}{2} ( 2+\ch_2) 
     \left ( \frac{\eh^2}{\ch^2} \tilde f_{BW}
     - \tilde f_{\Phi,1} \right) \right)
     \nonumber
  \\
             && \left.+ f_B \eh^2
                 \frac{\sh^2\ch_2^3}{2 \ch^2} \left( \tilde f_{\Phi,1} - 2
                \tilde \Delta_{4F} + \frac{\eh^2}{\sh^2} \tilde f_{BW}
                \right)
                + 4 \eh^2 \ch_2^3 f^{(3)}_{W^2\Phi^4} 
                \right ]\,
\nonumber
\end{eqnarray}
And the couplings to  $W^+_{\mu\nu} W^{- \, \nu\rho} V_\rho^{\; \mu}$ are: 
\begin{eqnarray}
  \lambda_\gamma =&& \frac{3}{2} \frac{\eh^2}{\sh^2}
                     \frac{\hat{M}_W^2}{\Lambda^2}  \Big[
                     {\left(f_{WWW} + \frac{\vh^2}{2\Lambda^2}
                     f^{(1)}_{W^3\Phi^2} \right)}
                     \nonumber
  \\
  && +
                     \frac{1}{2 \ch_2}
                     \frac{\vh^2}{\Lambda^2} f_{WWW} \left(  \eh^2 \tilde f_{BW}  -\left( 2
                     \tilde \Delta_{4F}   +  \tilde f_{\Phi,1}\right ) \ch^2 \right)\Big]
                    - \frac{\hat{M}_W^4}{2\Lambda^4} f^{(1)}_{W^2B\Phi^2}\;,
\nonumber
  \\
                  &&
\label{eq:lama}                     
\\
                     \lambda_Z =&&  \frac{3}{2} \frac{\eh^2}{\sh^2}
                     \frac{\hat{M}_W^2}{\Lambda^2}  \Big[{  \left(f_{WWW} + \frac{\vh^2}{2\Lambda^2} 
                                   f^{(1)}_{W^3\Phi^2} \right)}
                                   \nonumber
  \\
  &&-  
              \frac{\vh^2}{\Lambda^2} \frac{\sh^2(2+\ch_2)}{\sh_2 \sh_4}
              f_{WWW}\Big ( 4  \tilde \Delta_{4F} \ch^2 + 2  \tilde f_{\Phi,1} \ch^2
              -2 \eh^2\tilde f_{BW} \Big)
              \Big] + \frac{\hat{M}_W^4}{2\Lambda^4} \frac{\sh^2}{\ch^2} f^{(1)}_{W^2B\Phi^2} \;,
\nonumber
\end{eqnarray}
where $\hat{M}_W = \eh \vh/2\sh$.

\bibliography{references}

\end{document}